\documentclass{aa}  
\usepackage{graphicx}
\usepackage{txfonts}
\usepackage{hyperref}
\usepackage[dvipsnames]{xcolor}
\usepackage{booktabs}
\usepackage{soul}

\usepackage{siunitx}

\defcitealias{2024arXiv240501631B}{Paper\,I}
\defcitealias{2024arXiv240501634G}{Paper\,II}

\begin{document}

\title{JWST Imaging of the Closest Globular Clusters\,--\,III.\\ Multiple Populations along the low-mass Main Sequence stars of NGC\,6397}


\author{M.\,Scalco\inst{1,2}\fnmsep\thanks{\email{michele.scalco@inaf.it}}
\and   
M.\,Libralato\inst{2}
\and
R.\,Gerasimov\inst{3}
\and
L.\,R.\,Bedin\inst{2}
\and
E.\,Vesperini\inst{4}
\and
D.\,Nardiello\inst{5,2}
\and
A.\,Bellini\inst{6}
\and
M.\,Griggio\inst{1,2,6}
\and
D.\,Apai\inst{7,8}
\and
M.\,Salaris\inst{9}
\and
A.\,Burgasser\inst{10}
\and
J.\,Anderson\inst{6}
}

\institute{Dipartimento di Fisica e Scienze della Terra, Università di Ferrara, Via Giuseppe Saragat 1, Ferrara I-44122, Italy
\and
Istituto Nazionale di Astrofisica, Osservatorio Astronomico di Padova, Vicolo dell’Osservatorio 5, Padova I-35122, Italy
\and
Department of Physics and Astronomy, University of Notre Dame, Nieuwland Science Hall, Notre Dame, IN 46556, USA
\and
Department of Astronomy, Indiana University, Swain West, 727 E. 3rd Street, Bloomington, IN 47405, USA
\and
Dipartimento di Fisica e Astronomia "Galileo Galilei", Universit{\`a} di Padova, Vicolo dellOsservatorio 3, I-35122, Padova, Italy
\and
Space Telescope Science Institute, 3700 San Martin Drive, Baltimore, MD 21218, USA
\and
Department of Astronomy and Steward Observatory, The University of Arizona, 933 N. Cherry Avenue, Tucson, AZ 85721, USA
\and
Lunar and Planetary Laboratory, The University of Arizona, 1629 E. University Blvd., Tucson, AZ 85721, USA
\and
Astrophysics Research Institute, Liverpool John Moores University, 146 Brownlow Hill, Liverpool L3 5RF, UK
\and
Department of Astronomy \& Astrophysics, University of California, San Diego, La Jolla, California 92093, USA
}

\date{XXX,YYY,ZZZ}
 
\abstract
{Thanks to its exceptional near-infrared photometry, JWST can effectively contribute to the discovery, characterization, and understanding of multiple stellar populations in globular clusters, especially at low masses where the Hubble Space Telescope (HST) faces limitations. This paper continues the efforts of the JWST GO-1979 program in exploring the faintest members of the globular cluster NGC\,6397. Here we show that the combination of HST and JWST data allows us to identify two groups of MS stars (MSa, the first-generation, and MSb, the second-generation group). We measured the ratio between the two groups and combined it with measurements from the literature focused on more central fields and more massive stars compared to our study. Our findings suggest that the MSa and MSb stars are present in a $\approx$30-70 ratio regardless of the distance from the centre of the cluster and the mass of the stars used so far. However, considering the limited areal coverage of our study, a more comprehensive spatial analysis is necessary to definitively confirm complete spatial mixing.}

\keywords{techniques: photometric; catalogues; Hertzsprung-Russell and colour-magnitude diagrams; Population II; star clusters: individual: NGC\,6397}

\titlerunning{JWST Imaging of the Closest Globular Clusters\,--\,III.}
\authorrunning{M.\,Scalco et al.}
\maketitle

\section{Introduction}\label{sec:intro}

Many important questions in the study of globular clusters (GCs) benefit from the exquisite near-infrared (NIR) astrometry and photometry JWST can provide. An example is the phenomenon of the multiple stellar populations \citep[mPOPs; for a recent review, see][]{2019A&ARv..27....8G}.

Multiple stellar populations in GCs have been extensively studied for decades using the {\it Hubble Space Telescope} \citep[HST, since][]{2004ApJ...605L.125B}. The ultraviolet (UV) and optical photometry provided by HST are particularly effective in disentangling the mPOPs in colour-magnitude (CMD) and colour-colour diagrams \citep[e.g.][]{2017MNRAS.464.3636M}, spanning from the red-giant branch (RGB) to the main sequence (MS). However, as stars become fainter and redder, the effectiveness of UV filters diminishes. While optical filters maintain high photometric precision for observing MS stars, they alone may not always consent the identification of mPOPs as they lack the necessary information to differentiate stars with varying levels of oxygen, carbon, and nitrogen enhancement.

For these reasons, the analysis of mPOPs along the low-mass portion of the MS has predominantly relied on Near-Infrared (NIR) filters \citep[see][and reference therein]{2024arXiv240303262S} despite the inferior quality of HST's NIR data compared to UV/optical detectors. The new generation of detectors onboard JWST now provides high-resolution, deep NIR images, enabling continued exploration of mPOPs along the MS of GCs and potentially advancing it further, as indicated by preliminary studies \citep{2022MNRAS.517..484N,2023MNRAS.522.2429M,2023ApJ...953...62Z,2023A&A...679L..13C}.

This paper is the third of a series aimed at investigating one of the closest GCs to the Sun: NGC\,6397. NGC\,6397 is a close-by \citep[2.48~kpc;][]{2021BaumgardtGaiaDist}, metal-poor ($[\mathrm{Fe/H}]=-2.02$; \citealt{1996AJ....112.1487H,2010arXiv1012.3224H}) and old \citep[age$=$12.6~Gyr;][]{2018ApJ...864..147C} GC. The first paper of the series, \citet[hereafter \citetalias{2024arXiv240501631B}]{2024arXiv240501631B} used data from the Near Infrared Camera \citep[NIRCam,][]{2023PASP..135b8001R} onboard JWST to characterize the white dwarf (WD) cooling sequence (CS) of this cluster, while the second paper, \citet[hereafter \citetalias{2024arXiv240501634G}]{2024arXiv240501634G} is focused on the brown dwarfs (BDs) characterization of NGC\,6397.  

The presence of mPOPs in this stellar system was initially explored by \citet{2012ApJ...745...27M} using data obtained with HST. \citet{2012ApJ...745...27M} found that the MS of the cluster splits into two distinct components. One component (namely MSa) consists of a primordial population, resembling the composition of field stars. The other component (namely MSb) corresponds to a second generation of stars characterized by enhanced sodium and nitrogen levels, reduced carbon and oxygen content, and a slightly increased helium abundance. The authors also found that MSb stars are more numerous ($\sim$70\% of the total) than MSa ($\sim$30\%). Here we make use of the astro-photometric catalogue of \citetalias{2024arXiv240501631B} to analyze the mPOPs in the lower MS of NGC\,6397.

The paper is organized as follows. Section~\ref{sec:obs} summarizes the data set and reduction; Section~\ref{sec:mpops} describes the identification of the mPOPs, the isochrone fit and the radial and mass distribution of the two populations. Conclusions are presented in Section~\ref{sec:conc}.
    
\section{Observations and data reduction}   
\label{sec:obs}   

The data were obtained as part of three programs:

\begin{itemize}
\item HST GO-10424 (PI: Richer; taken in $\sim$2005.2). This set comprises images (with a combination of long and short exposure times) taken with Wide Field Channel (WFC) of the Advanced Camera for Surveys (ACS) in F606W and F814W filters;
\item HST GO-11633 (PI: Rich; taken in $\sim$2010.2). As before, this data set includes both long and short exposures in ACS/WFC F606W and F814W filters.
\item JWST GO-1979 (PI: Bedin; taken in $\sim$2023.2). The JWST data were obtained with the NIRCam camera in F150W2 and F322W2 filters.
\end{itemize}

\begin{figure}
\includegraphics[width=\columnwidth]{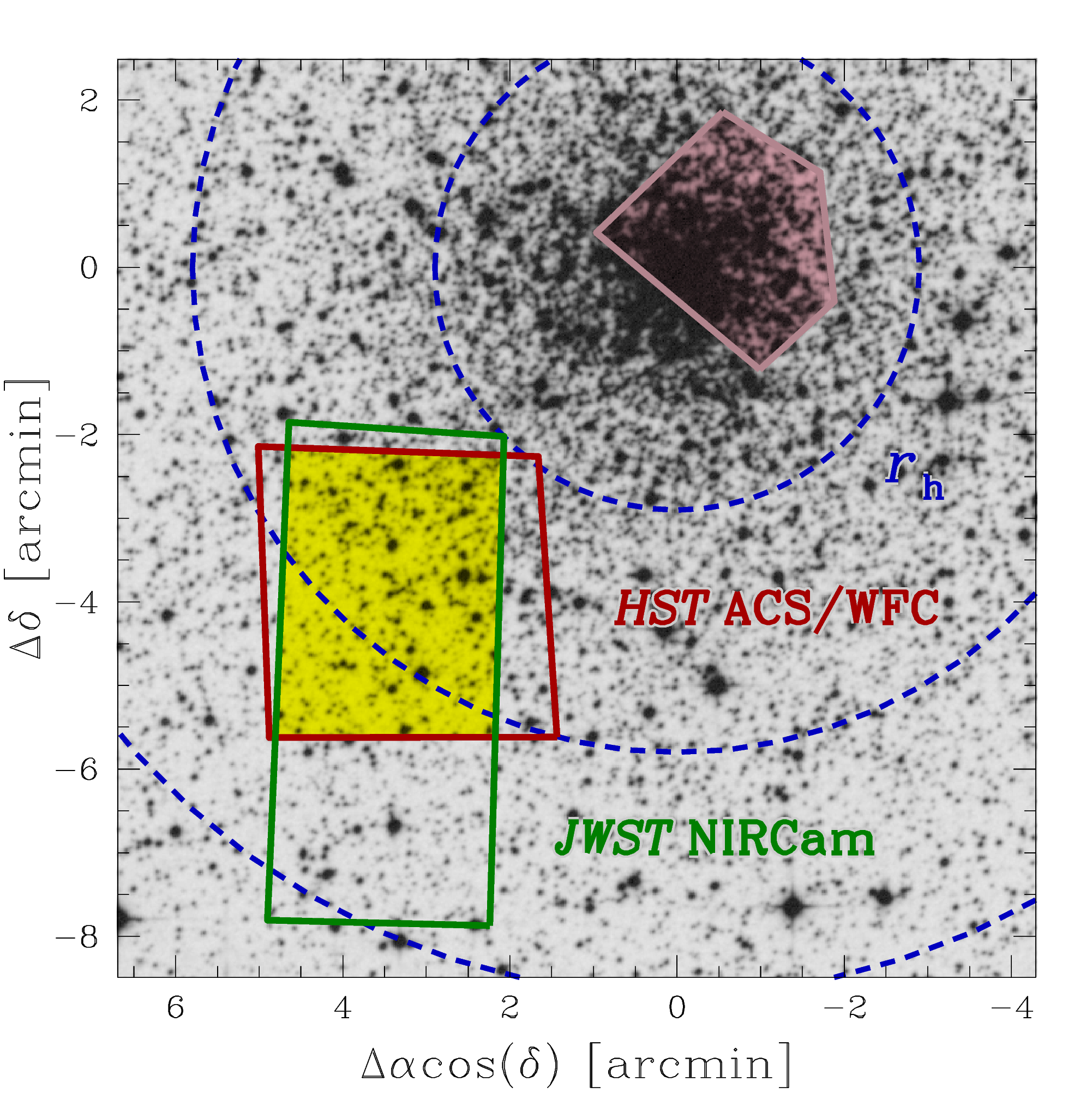}
\caption{Outlines of the ACS/WFC HST field observed under the GO-10424 and GO-11633 programs (in red) and the NIRCam JWST field observed under the GO-1979 program (in green), superimposed on a DSS image of NGC\,6397. The region where the two data sets overlap is highlighted in yellow. We also show, in magenta, the field analysed in \citet{2012ApJ...745...27M}. Units are in arcmin measured from the centre of the cluster. The blue dashed circles mark the half-light radius ($r_{\rm h}=2^\prime_{\cdot}90$; \citealt{1996AJ....112.1487H,2010arXiv1012.3224H}), $2\,r_{\rm h}$ and $3\,r_{\rm h}$.}
\label{fig:FOV}
\end{figure}

Figure\,\ref{fig:FOV} shows the position of our datasets in the field-of-view (FOV) overlaid on a Digital Sky Survey\footnote{\href{https://archive.eso.org/dss/dss}{https://archive.eso.org/dss/dss}} (DSS) image of NGC\,6397.

We made use of calibrated but unresampled exposures (\texttt{\_flc} and \texttt{\_cal} fits files for HST and JWST, respectively). The data reduction was performed as described in \citetalias{2024arXiv240501631B}, i.e., a combination of first- and second-pass photometry for HST \citep[see][for details]{2017ApJ...842....6B,2018ApJ...853...86B,2018MNRAS.481.3382N,2018ApJ...854...45L,2022LibralatoPMcat,2021MNRAS.505.3549S} and JWST images \citep[see][for details]{2022MNRAS.517..484N,2023MNRAS.525.2585N,2023AN....34430006G,2023LibralatoNIRISS,2024LibralatoMIRI}. The astrometry was anchored to an absolute reference-frame system provided by the {\it Gaia} Data Release 3 (DR3) catalogue \citep{2016A&A...595A...1G,2023A&A...674A...1G}, and the photometry was calibrated to the VEGA-magnitude photometric system. Proper motions (PMs) were calculated by measuring the displacements between the JWST observations and the earliest HST observations, covering a temporal baseline of $\sim$18 years. Detailed descriptions of the data sets and reduction are provided in \citetalias{2024arXiv240501631B}.

The complete catalogue contains spurious detections such as cosmic rays or noise peaks in individual images. For this work, only a subset of well-measured cluster members was selected using the procedure outlined in Appendix~A of \citetalias{2024arXiv240501631B}. This section process is based on diagnostic parameters such as the quality-of-PSF fit parameter (\texttt{QFIT}), the excess/deficiency of the source's flux \citep[\texttt{RADXS}, see][]{2008ApJ...678.1279B}, the local sky noise (\texttt{rmsSKY}), and the utilization of PM-based memberships criteria.

\section{The split of the main sequence in NGC\,6397 at lower stellar masses}\label{sec:mpops}

\citet{2012ApJ...745...27M} focused on the upper part of the MS within the mass range of $\sim$0.56 to $\sim$0.62~M$_\odot$, covering the central region of the cluster spanning from the centre to $\sim$2~arcmin (equivalent to $\sim$0.7$r_{\rm h}$). In this study, we used our HST and JWST datasets to investigate the two populations in the lower part of the MS, with masses ranging from $\sim$0.09 to $\sim$0.18~M$_\odot$, in an outer region spanning from $\sim$3.2 to $\sim$7~arcmin (equivalent to $\sim$1.1$r_{\rm h}$ to $\sim$2.4$r_{\rm h}$) of NGC\,6397.

We followed the methodology outlined by \citet{2012A&A...540A..16M} \citep[see also][]{2007AJ....133.1658S,2017ApJ...842....7B} to correct our photometry for differential reddening and photometric zero-point variations across the FOV. Then, we made use of two-pseudo-colour diagrams \citep[TpCDs, see][]{2015MNRAS.447..927M,2015ApJ...808...51M} to identify the two groups of stars in the MS of NGC\,6397. 

\begin{figure*}
\includegraphics[width=\textwidth]{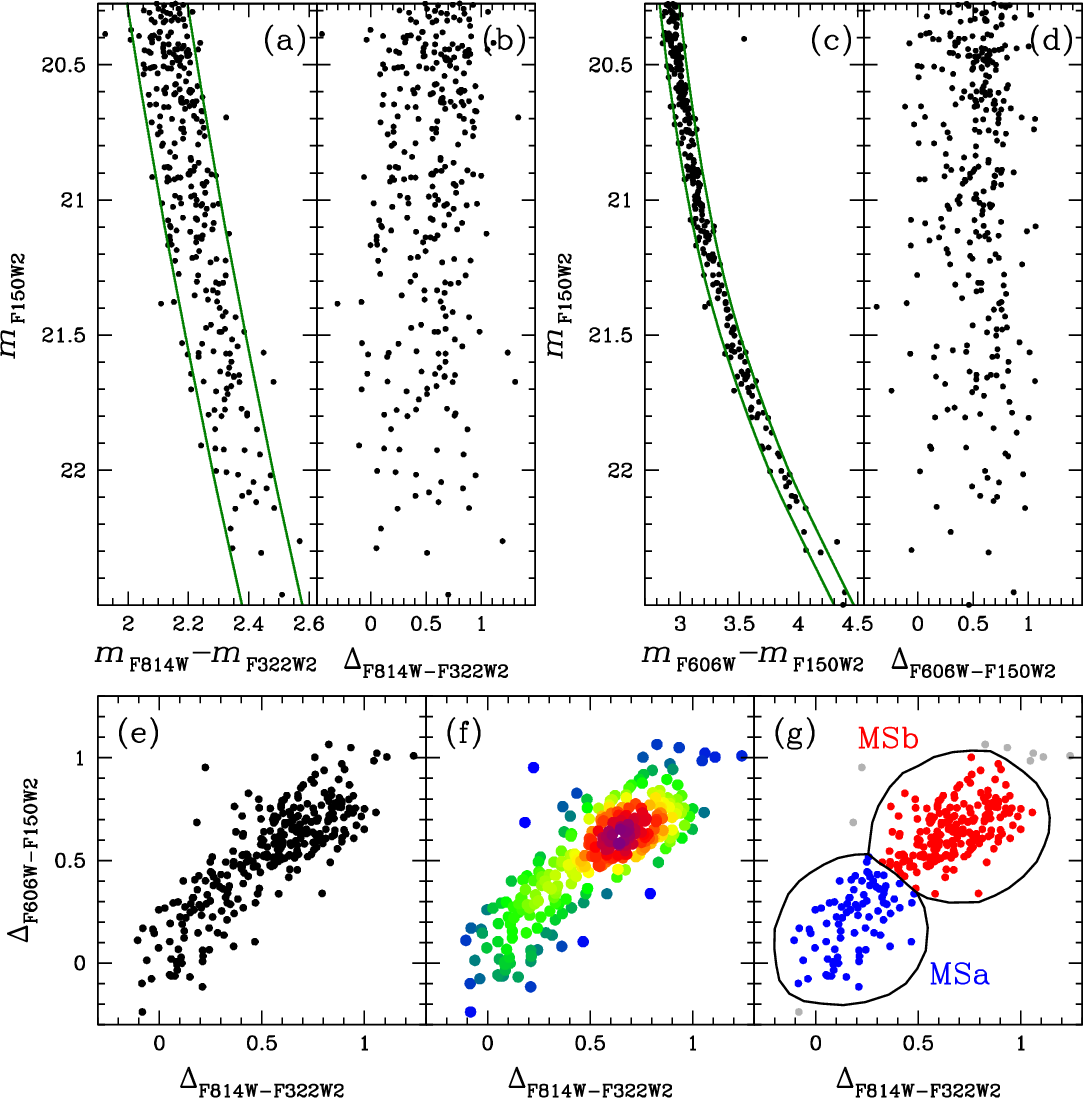}
\caption{Procedure used to identify the two sequences of stars in the lower part of the MS of NGC\,6397. (a) $m_{\rm F150W2}$ versus $m_{\rm F814W}-m_{\rm F322W2}$ CMD of the stars in our sample, corrected for differential reddening and photometric zero-point variations. The two green lines (drawn by hand) are used to rectify the diagram. (b) Rectified $m_{\rm F150W2}$ versus $\Delta_{\rm F814W-F322W2}$ CMD. (c)-(d) Same as (a)-(b) but for the $m_{\rm F150W2}$ versus $m_{\rm F606W}-m_{\rm F150W2}$ CMD. (e) $\Delta_{\rm F606W-F150W2}$ versus $\Delta_{\rm F814W-F322W2}$ TpCD. (f) Hess diagram of the TpCD. (g) The two stellar populations identified in the TpCD (within the black envelopes), MSa (in blue) and MSb (in red). Stars outside the black envelopes are represented in grey.}
\label{fig:ChM}
\end{figure*}

Panels\,(a) and (c) of Fig.\,\ref{fig:ChM} present the differential-reddening-and-zero-point-corrected $m_{\rm F150W2}$ versus $m_{\rm F814W}-m_{\rm F322W2}$ and $m_{\rm F150W2}$ versus $m_{\rm F606W}-m_{\rm F150W2}$ CMDs, respectively. In each CMD, we drew by hand two fiducial curves enclosing the MS (represented in green in panels\,(a) and (c)), that were used to rectify the sequence. The rectified CMDs are presented in panels\,(b) and (d), where $\Delta_{\rm F814W-F322W2}$ and $\Delta_{\rm F606W-F150W2}$ denote the $m_{\rm F814W}-m_{\rm F322W2}$ and $m_{\rm F606W}-m_{\rm F150W2}$ rectified and normalised pseudo-colours, respectively. Two sequences are faintly discernible in these diagrams, with one more populated on the red side of the plot and another less populated on the blue side of the plot.

We combined the two verticalized diagrams to construct the TpCD (panel\,e). Panel\,(f) displays the Hess diagram of the TpCD, colour-coded according to the local density of points. The colour goes from blue (lowest density) to green (average density) and red (highest density). Two components are visible: a clump around (0.65,0.60) representing the bump of MSb stars, and a tail extending towards the lower left part of the plot representing the MSa stars.

The black envelopes (hand-defined) in panel\,(g) tentatively delineate the two components in the TpCD, with MSa stars in blue and MSb stars in red. Stars outside the envelopes are rejected and represented in grey. We note that this represents an initial exploration, and a more detailed quantitative analysis will follow. The MSa component comprises 84 stars, representing the 29\%$\pm$4\% of the total, while the MSb component comprises 201 stars, representing the 71\%$\pm$6\% of the total. The errors in this and subsequent fractions are estimated by linear propagation of Poisson noise. These values are consistent with those reported in \citet[$\sim$30\%$\pm$3\% for the MSa and $\sim$70\%$\pm$3\% for the MSb]{2012ApJ...745...27M}. 

\begin{figure*}
\includegraphics[width=\textwidth]{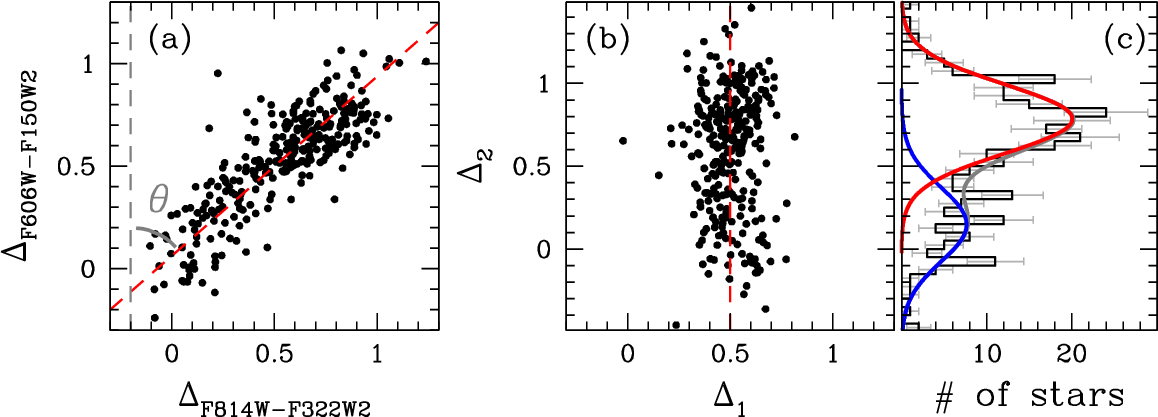}
\caption{(a) Same TpCD presented in panels\,(e) of Fig.\,\ref{fig:ChM}. The red dashed line defines an angle $\theta$ with the y-axis. (b) $\Delta_{2}$ versus $\Delta_{1}$ diagram obtained by the counterclockwise rotation of panel\,(a) by an angle $\theta$ around the point (0.5,0.5). (c) Histogram of the $\Delta_{2}$ distribution with error bars representing Poisson errors. The best-fitting double-Gaussian function is represented in grey, while the two components are plotted in blue and red.}
\label{fig:Gaus}
\end{figure*}

To better estimate the number of stars in each group while considering potential contamination between the populations, we adopted an alternative approach. Panel\,(a) of Fig.\,\ref{fig:Gaus} displays the same TpCD as presented in panels\,(e) of Fig.\,\ref{fig:ChM}. Initially, we fitted a straight line to the barycenters of the two populations in the TpCD (red dashed line), defining an angle $\theta$ with respect to the y-axis of the plot. Panel\,(b) shows the $\Delta_{2}$ versus $\Delta_{1}$ diagram obtained by a counterclockwise rotation of the TpCD by an angle $\theta$ around the point (0.5,0.5) so as to have the red line parallel to the y-axis. Finally, panel\,(c) presents the histogram of the $\Delta_{2}$ distribution, with error bars representing Poisson errors. We fitted a double-Gaussian function to the histogram (depicted in grey), with the two components associated with MSa and MSb represented in blue and red, respectively. Once again, we determined that MSa and MSb comprise 29\%$\pm$4\% and 71\%$\pm$6\% of the total, respectively. These values are in agreement with those obtained previously.

\begin{figure*}
\includegraphics[width=\textwidth]{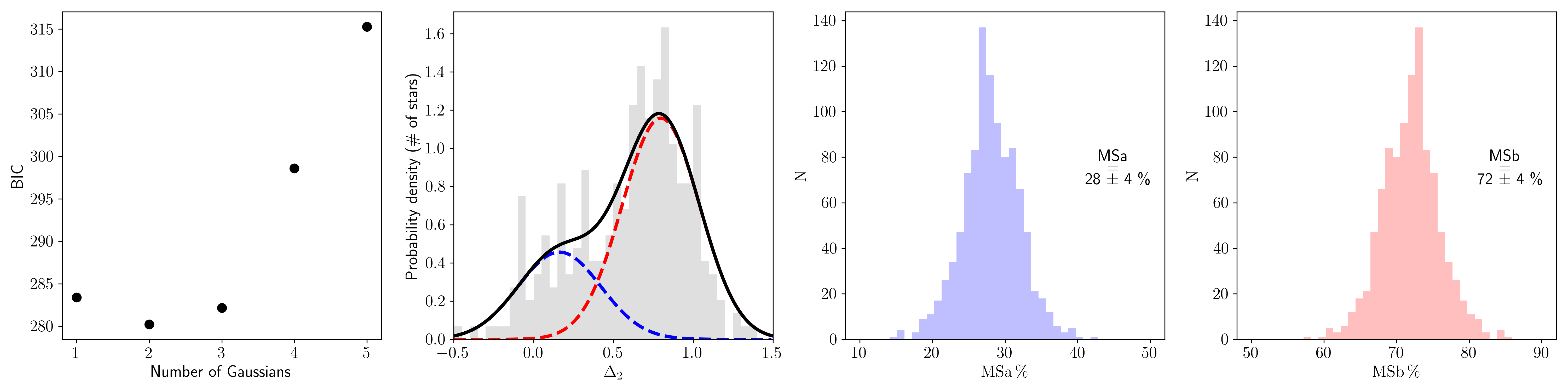}
\caption{Analysis of the $\Delta_2$ distribution using GMM. \textit{First panel}: BIC as a function of the number of Gaussian components considered in the GMM. \textit{Second panel}: histogram of the $\Delta_2$ distribution fitted with a two-Gaussian model (black continuous line). The two components of the two-Gaussian model are plotted as blue and red dashed lines. \textit{Third and fourth panels}: distribution of the ratio of the two populations obtained from 1000 repetitions of the GMM fitting process. The reported values represent the median and standard deviation of the distribution.}
\label{fig:GMM}
\end{figure*}

We assessed the robustness of our fitting method using a Gaussian mixture model (GMM) on the $\Delta_2$ star distribution, employing the expectation-maximization algorithm from the \texttt{scikit-learn} package \citep{scikit-learn}. To determine the optimal number of Gaussian components for the $\Delta_2$ distribution, we calculated the Bayesian Information Criterion (BIC) for models with 1 to 5 Gaussian components. The first panel of Fig.\,\ref{fig:GMM} shows the BIC as a function of the number of Gaussian components, with the two-Gaussian model having the lowest BIC. The second panel of Fig.\,\ref{fig:GMM} presents the histogram of the $\Delta_2$ distribution fitted with the two-Gaussian model. To further ensure robustness, we repeated the GMM fitting process 1000 times. In each iteration, we randomly selected 300 stars from our sample and calculated the fraction of stars in each MS. The third and fourth panels of Fig.\,\ref{fig:GMM} display the distribution of the ratio of the two populations from these 1000 iterations. The reported values are the median and standard deviation of the distribution, which are consistent with the values obtained from the analysis discussed earlier.

\subsection{Isochrone fit}

\begin{figure}
\includegraphics[width=\columnwidth]{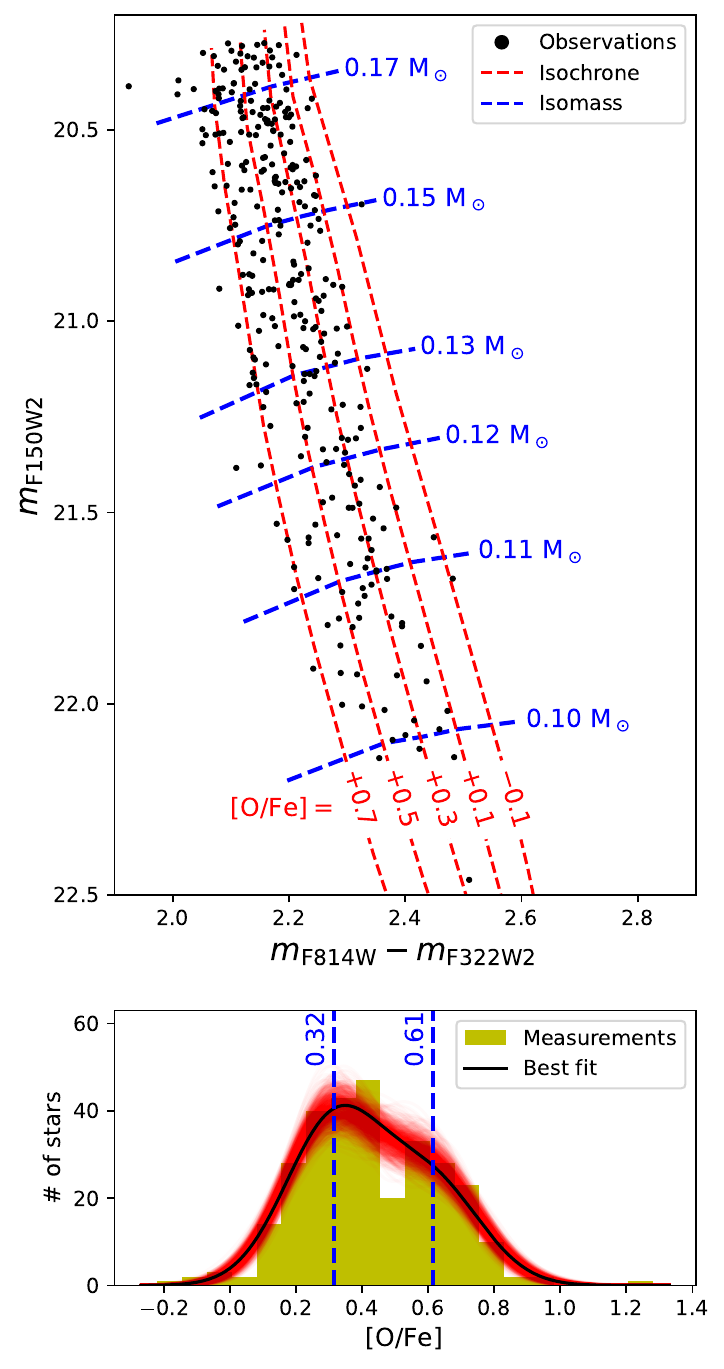}
\caption{\textit{Top:} model isochrones calculated in this study overplotted on NIRCam photometry of NGC\,6397. The isochrones are labelled by the adopted oxygen abundance. Linearly interpolated isomass lines are shown for selected masses. \textit{Bottom:} histogram of measured $[\mathrm{O/Fe}]$ values for the individual stars in the cluster. The best-fit two-component Gaussian mixture is overplotted, with the central values of both components indicated with vertical lines. The uncertainty in the best-fit distribution was estimated using $1000$ bootstrap trials, shown with semi-transparent red lines.}
\label{fig:mPOPs}
\end{figure}

At the range of effective temperatures considered in this study (lower MS below the MS knee, $T_\mathrm{eff}=3000-\qty{4000}{\K}$), broadband photometry is primarily influenced by the abundances of oxygen and titanium due to their role in molecular opacity \citep{roman_47tuc}. Since the member-to-member scatter in $[\mathrm{O/Fe}]$ is the primary tracer of mPOPs in GCs, the individual populations can be identified in the CMD through a careful selection of photometric bands. The F322W2 band of NIRCam is particularly sensitive to the oxygen abundance due to its proximity to the fundamental vibrational bands of water near \qty{3}{\micro\meter} \citep{2001A&A...372..249J}. The second band is most well-placed in the optical regime where $\mathrm{H_2O}$ absorption is minimal; however, wavelengths redder than $\sim\qty{0.7}{\micro\meter}$ are preferred to avoid the strongest electronic bands of $\mathrm{TiO}$ \citep{1998A&A...337..495P}. Of the filters considered in this work, the F814W band of ACS is therefore the most suitable. The oxygen content of individual stars in NGC\,6397 can be inferred from the $m_{\rm F814W}-m_{\rm F322W2}$ colour with minimal systematic errors due to potentially unknown abundances of other elements.

To derive $[\mathrm{O/Fe}]$ from photometry, we first chose a model isochrone from the \texttt{SANDee} grid (\citetalias{2024arXiv240501634G}; Alvarado et al. \textit{in prep}) that provides the best representation of the $m_{\rm F150W2}$ versus $m_{\rm F814W}-m_{\rm F322W2}$ and $m_{\rm F150W2}$ versus $m_{\rm F606W}-m_{\rm F150W2}$ CMDs. The chosen isochrone had $[\mathrm{Fe/H}]=-1.75$ and $[\alpha/\mathrm{Fe}]=0.3$. We then calculated $5$ new model isochrones, varying $[\mathrm{O/Fe}]$ from $-0.1$ to $0.7$ in steps of $0.2$. The oxygen abundance was updated in the model atmospheres of the best-fit \texttt{SANDee} isochrone with $\log(g)\ge4$, as well as the corresponding evolutionary models and boundary condition tables, following \citet{roman_47tuc,roman_omegacen}. For all new isochrones, we adopted the age of $\qty{12.6}{Gyr}$ from \citet{2018ApJ...864..147C}, the distance of $\qty{2.458}{kpc}$ from \citet{2021BaumgardtGaiaDist} and the optical reddening of $E(B-V)=0.18$ from \citet{2003A&A...408..529G}. The isochrones are overplotted on the observed CMD in the upper panel of Figure~\ref{fig:mPOPs}.

Model isochrones for other oxygen abundances were obtained by linearly interpolating or extrapolating the mass-color and mass-magnitude relationships. The values of $[\mathrm{O/Fe}]$ for individual stars were then determined by searching for the interpolated isochrones that pass through the observed colours and magnitudes using Brent's method \citep{1973SJNA...10..327B}. Overall, the oxygen abundance of $\sim300$ members of NGC\,6397 was measured. The distribution of these measurements is shown in the lower panel of Figure~\ref{fig:mPOPs}. The modal values of $[\mathrm{O/Fe}]$ for the two populations in the cluster were taken as the central values of the Gaussian components in a two-component Gaussian mixture fit to the inferred distribution. The fit was carried out using the expectation-maximization algorithm implemented in the \texttt{scikit-learn} package. The two-component fit is shown in the lower panel of Figure~\ref{fig:mPOPs}. The modal values of $[\mathrm{O/Fe}]$ were estimated as $0.32\pm0.02$ and $0.61\pm0.02$ for the two populations. We also estimated the expected systematic offsets in our $[\mathrm{O/Fe}]$ measurements by comparing other possible choices for the best-fit \texttt{SANDee} isochrone to the isochrones used in this work. The results of this analysis are summarized in Table~\ref{tab:systematics_in_O}.

\begin{center}
\begin{table}[t]%
\centering
\caption{Effect of choosing a different base model isochrone on the oxygen abundances of individual stars.\label{tab:systematics_in_O}}%
\tabcolsep=0pt%
\begin{tabular*}{20pc}{@{\extracolsep\fill}cc@{\extracolsep\fill}}
\toprule
\textbf{Change in base isochrone} & \textbf{Expected offset in $[\mathrm{O/Fe}]$}  \\
\midrule
Increase $[\alpha/\mathrm{Fe}]$ by $0.2$ & $+0.18$    \\
Decrease $[\mathrm{Fe/H}]$ by $0.2$ & $-0.12$    \\
Increase $E(B-V)$ by $0.02$ & $+0.09$    \\
Decrease age by $\qty{2}{Gyr}$ & $-0.002$    \\
\bottomrule
\end{tabular*}
\end{table}
\end{center}

In \citet{2012ApJ...745...27M}, the representative $[\mathrm{O/Fe}]$ of the two populations were estimated as $0.1$ and $0.45$ based on the differences between observed colours and model atmospheres. The values found in this work are $\sim0.2$ higher. 
This discrepancy can be reconciled only by reducing the metallicity of the best-fit isochrone by $\sim0.3-0.4$, i.e., by setting it to $[\mathrm{Fe/H}]\sim-2$. The implied lower metallicity would in fact be more consistent with other photometric \citep{2018ApJ...864..147C} and spectroscopic \citep{NGC6397_MUSE} estimates in the literature. The higher metallicity of NGC\,6397 computed in this work near the end of the MS is likely due to imperfectly modelled physical processes in low-temperature stellar atmospheres such as depletion of gas-phase elements onto dust grains or non-equilibrium chemistry. A similar effect has been observed in a different NIRCam field of the cluster (\citetalias{2024arXiv240501634G}).

The relative difference in $[\mathrm{O/Fe}]$ between the two populations is approximately equal to that determined by \citet{2012ApJ...745...27M}, confirming the correspondence between the cluster populations identified in both studies.

\subsection{Radial and mass distribution of the two populations}
Although the fractions of the two populations obtained in our study generally align with those reported by \citet{2012ApJ...745...27M}, we subdivided our sample to investigate potential dependencies of these populations on the radial distance from the cluster's centre and stellar mass. To do this, we considered the two groups of stars defined in panel\,(g) of Fig.\,\ref{fig:ChM}.

\begin{figure}
\includegraphics[width=\columnwidth]{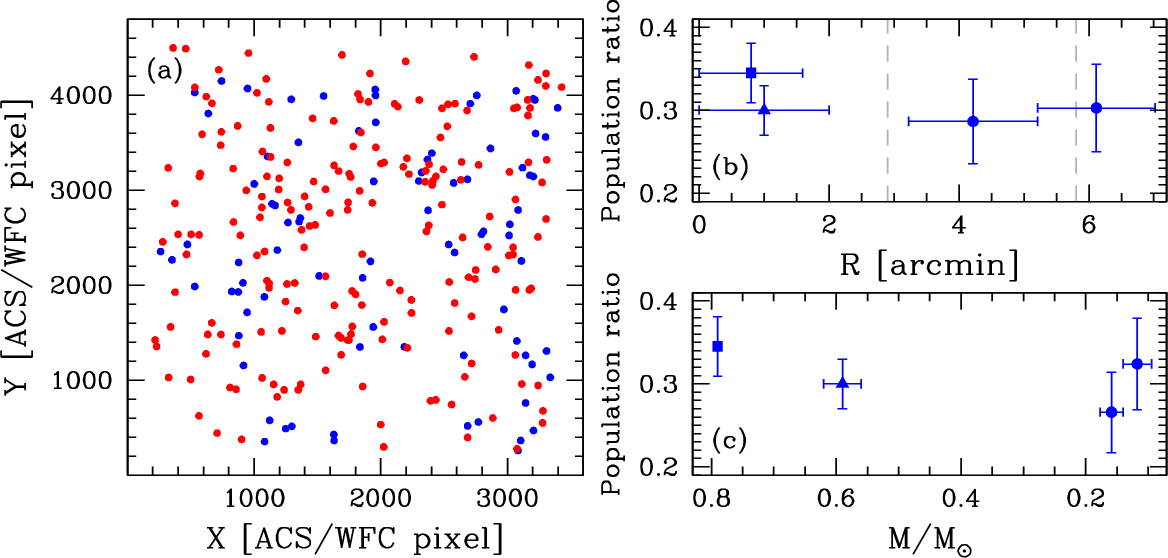}
\caption{(a) Spatial distribution of the two populations in the X-Y plane. Points are colour-coded as in panel\,(g) of Fig.\,\ref{fig:ChM}. (b) Fraction of MSa stars as a function of radial distance from the centre of the cluster, R. The values obtained in this work are represented with circles, while the estimates from \citet{2012ApJ...745...27M} and \citet{2017MNRAS.464.3636M} are shown with triangles and squares, respectively. Each dataset corresponds to different mass ranges (see panel\,(c)). The grey dashed lines are set at $r_{\rm h}$ and $2r_{\rm h}$. (c) Fraction of MSa stars as a function of stellar mass. Symbols have the same meaning as in panel\,(b).}
\label{fig:Figure2}
\end{figure}

Panel\,(a) of Fig.\,\ref{fig:Figure2} illustrates the spatial distribution of the two stellar populations in the X-Y plane. Points are colour-coded as in panel\,(g) of Fig.\,\ref{fig:ChM}. The two populations show no clear difference in their spatial distributions.

We examined the radial profile of the relative number of the two populations by dividing the stars into two equally-populated radial bins. We then evaluated for each bin the ratio of the MSa population relative to the total number of stars in that bin. Panel\,(b) of Fig.\,\ref{fig:Figure2} shows our measurements (filled circles) together with the values reported by \citet[triangles]{2012ApJ...745...27M} and \citet[squares]{2017MNRAS.464.3636M} for the central part of the cluster. It is important to note that each dataset corresponds to different mass ranges (see discussion below). The combined dataset spans a radial distance from the centre of the cluster out to $\sim$7~arcmin, corresponding to $\sim$2.4$r_{\rm h}$. This comprehensive analysis demonstrates that the two populations of NGC\,6397 have reached complete spatial mixing. NGC\,6397 has a half-mass relaxation time of t$_{r{\rm h}}\sim$0.5~Gyr \citep{2021MNRAS.505.5978V} and is thus a dynamically old cluster \citep[age/t$_{r{\rm h}}\sim$25; see also][]{2009MNRAS.395.1173G}; various investigations studying the dynamical evolution of multiple-population clusters \citep[see e.g.][]{2013MNRAS.429.1913V,2021MNRAS.502.4290V,2019ApJ...884L..24D} predict that in clusters with this dynamical age the two populations should indeed be completely mixed and have lost memory of any initial spatial differences.

Finally, panel\,(c) presents the population ratio as a function of stellar mass. The average mass of the stars was inferred by means of the isochrones presented in Fig.\,\ref{fig:mPOPs}. We divided our sample into two equally populated mass bins and evaluated the ratio of the MSa population in each bin. Our estimates are represented in panel\,(c) along with the values reported by \citet{2012ApJ...745...27M} and \citet{2017MNRAS.464.3636M} using the same symbols as in panel\,(b). Similarly to what was found in the analysis of the radial distribution of the two populations, we find that the fraction of stars in the two populations is also independent of the stellar mass. This result is consistent, in general, with what was found for other clusters \citep[e.g.][]{2019MNRAS.484.4046M,2022ApJ...927..207D,2023MNRAS.522.2429M,2023ApJ...953...62Z}. Since in our analysis different mass intervals are studied in different radial regions we can not carry out a consistent analysis of the stellar mass function of the two populations and its radial variation \citep[see e.g.][for a study of mass segregation and radial variation of the stellar mass function in this cluster]{1995ApJ...452L..33K}.

\section{Conclusions}\label{sec:conc}
This paper represents the third instalment in a series, with a focus on the characterization of the mPOPs in NIR within the lower MS of NGC\,6397 using high-precision photometry with HST and JWST. The TpCD highlights the presence of two components, namely MSa and MSb, associated with a primordial and second-generation population, respectively.

MSa and MSb are present in a $\approx$30-70 ratio, consistent with findings from previous studies of this GC focused on its centermost region. Combining the results in this work with those from the literature, we have found that the MSa/MSb ratio does not vary with the distance from the centre of the cluster; NGC\,6397 is a dynamically old cluster and this finding is consistent with the results of numerical studies predicting that initial differences between the spatial distributions of mPOPs are gradually erased during the cluster long-term evolution and expected to be completely lost in the dynamically older clusters. However, due to the limited areal coverage of our study, this result should be interpreted with caution. A more extensive spatial coverage would be necessary to confirm complete spatial mixing definitively.

The JWST near-infrared photometry data analysed in this paper, allowed us to probe and identify mPOPs in a low-mass interval ($\sim 0.09-0.18$ M$_{\odot}$) never studied before for this cluster. A comparison of our data with those of studies covering stellar mass intervals shows that the mass fraction is also independent of the stellar mass. This is consistent with what was found for other clusters \citep[e.g.][]{2019MNRAS.484.4046M,2022ApJ...927..207D,2023MNRAS.522.2429M,2023ApJ...953...62Z} and could be used to put additional constraints on mPOP formation scenarios \citep[see discussion in, e.g.,][]{2023MNRAS.522.2429M}. Note that each of the three radial intervals analyzed for this cluster covers different mass intervals; the population ratio for different mass intervals is thus calculated at different radial distances; while we cannot a priori exclude the possibility that there are radial and mass dependencies cancelling each other and resulting in a population ratio independent of mass and radius, this would be an unlikely coincidence. For a comprehensive investigation of the stellar mass function and its radial variation, future observations providing data covering the same mass range at different radial distances will be necessary.

\begin{acknowledgements}
Michele Scalco and Luigi Rolly Bedin acknowledge support by MIUR under the PRIN-2017 programme \#2017Z2HSMF, and by INAF under the PRIN-2019 programme \#10-Bedin. LRB, DN, MG and MSc acknowledge support by INAF under the WFAP project, f.o.:1.05.23.05.05. EV acknowledges support from NSF grant AST-2009193.
\end{acknowledgements}

\bibliographystyle{aa}
\bibliography{biblio.bib}

\end{document}